\documentclass[]{JHEP3}

\JHEPspecialurl{http://jhep.sissa.it/JOURNAL/JHEP3.tar.gz}

\usepackage{graphicx}
\usepackage{amsmath}
\usepackage{epsfig,multicol,bbm}

\newcommand{\be}{\begin{equation}}
\newcommand{\ee}{\end{equation}}
\newcommand{\bea}{\begin{eqnarray}}
\newcommand{\eea}{\end{eqnarray}}

\newcommand{\mc}{\mathcal}
\newcommand{\gsim}{\gtrsim}

\newcommand{\beqa}{\begin{eqnarray}}
\newcommand{\eeqa}{\end{eqnarray}}

\newcommand{\vo}{\mathcal{V}}

\title{Moduli destabilization via gravitational collapse}

\author{ Dong-il Hwang\\
Center for Quantum Spacetime, Sogang University, Seoul 121-742, Republic of Korea\\
Institute of Basic Science, Sogang University, Seoul 121-742, Republic of Korea\\
 E-mail: \email{dongil.j.hwang@gmail.com}}

\author{Francisco G. Pedro\\
	Deutsches Elektronen-Synchrotron DESY, Theory Group, 22607 Hamburg, Germany\\
	E-mail:  \email{francisco.pedro@desy.de}}

 \author{Dong-han Yeom\\
 Center for Quantum Spacetime, Sogang University, Seoul 121-742, Republic of Korea\\
 Yukawa Institute for Theoretical Physics, Kyoto University, Kyoto 606-8502, Japan\\
 E-mail: \email{innocent.yeom@gmail.com}}

\abstract{
We examine the interplay between gravitational collapse and moduli stability in the context of black hole formation.
We perform numerical simulations of the collapse using  the double null formalism and show that the very dense regions one expects to find in the process of black hole formation are able to destabilize the volume modulus. We establish that the effects of the destabilization will be visible to an observer at infinity, opening up a window to a region in spacetime where standard model's couplings and masses can differ significantly from their background values.
}

\preprint{YITP-13-43, DESY 13-105}

\begin{document}

\section{Introduction}

One of the most striking features of string theory compactifications is the close connection between the geometry of the compact space and the four dimensional physics. While in principle one is free to choose the space in which one compactifies the higher dimensional theory, this connection forces the choice of geometries that yield the desired four dimensional physics. For a variety of reasons supersymmetry is a highly desirable feature to have in the four dimensional theory.  For compactifications of ten dimensional type IIB string theory, the requirement of $\mc{N}=1$ theories in 4D forces one to consider Calabi-Yau orientifold compactifications \cite{Grimm:2004uq, Giddings:2001yu}.

The geometry of these Calabi-Yau spaces is parameterized by the geometric moduli: K\"ahler and complex structure. Intuitively, K\"ahler moduli describe the volumes while complex structure moduli give the shape of the compactification space. From the four dimensional theory point of view they are Planck coupled massless scalar fields. Such fields face severe constraints both from fifth-force constraints \cite{Adelberger:2003zx} and from consistent cosmological evolution. Furthermore the vacuum expectation values of these fields determine the masses and couplings of the four dimensional field theory, and so unstabilized moduli lead to ill defined physical spectrum. We then see that in order to have a sensible theory of cosmology and particle physics in four dimensions it is imperative that these flat directions are lifted. 

Over the last decade significant progress has been made in the physics of moduli stabilization, in particular in type IIB string theory, where a combination of gauge flux in the extra dimensions and perturbative and non-perturbative corrections to the tree level effective action have yielded quasi-realistic compactifications of the ten dimensional theory \cite{Giddings:2001yu,Becker:2002nn,Kachru:2003aw,Balasubramanian:2005zx}.
Of particular interest are the LARGE volume compactifications of \cite{Balasubramanian:2005zx} due to their robustness and rich phenomenology (for a review of this scenario see e.g. \cite{Conlon:2006gv,Cicoli:2009zh}). These compactifications allow for the stabilization of the compact space at a non-supersymmetric AdS minimum at exponentially large volumes, allowing us to have control over the perturbative expansion without having to pay for it with fine-tuning of the parameters.

It is usually assumed that moduli stabilization happens in the same way throughout spacetime, however such assumption needs to be checked. Given that the moduli vevs determine the masses and couplings of the particles in the four dimensional theory, these particles will source the moduli potential and can in principle distort it, shifting the vevs of the moduli fields.
The robustness of the moduli stabilization mechanism against local perturbations sourced by matter fields has been studied in \cite{Conlon:2010jq} where it was found that even the densest known forms of matter could not have a measurable effect in the potential for the lightest Plank coupled modulus. The fundamental reason for this was that even within the densest astrophysical objects, like neutron stars, there is a large hierarchy between the scale of the modulus potential and the scale of the local perturbation: $\Lambda_{\mathrm{matter}}/M_{\mathrm{P}} \ll 1$. There were however two notable exceptions to this behaviour in the context of systems undergoing gravitational collapse: the superinflationary expansion of the compact space at the final stages of the collapse of a positively curved matter dominated FRW universe and decompactification in the process of black hole formation. For other setups in which a localized distribution of matter distorts the potential of a gravitationally coupled scalar field see also \cite{Khoury:2003rn,Green:2006nv}.

In this paper we will analyze the interplay between moduli stabilization and gravitational collapse in the formation of a black hole.
This constitutes an extension of the work carried out in \cite{Conlon:2010jq} and a check of the results reported there. The method used in \cite{Conlon:2010jq} for the study of gravitational collapse consisted in gluing a positively curved FRW universe filled with matter to an exterior Schwarzschild spacetime, this allowed for a study of the collapse \`a la Oppenheimer-Snyder \cite{Oppenheimer:1939ue}.

This previous work should be extended in two ways. Firstly, the Oppenheimer-Snyder collapse relies on a junction between a Schwarzschild black hole and a FRW universe permeated by a perfect fluid. This setup seems to be quite idealized and so we aim to extend the analysis for a more generic geometry and initial conditions using a dynamical metric and a dynamical matter field. Secondly, we need to study the causal structure during the gravitational collapses to understand whether the destabilized effect can affect the future infinity or if it is inside of the event horizon and hence there is no hope to see any effects of the destabilization; whether the destabilized field is maintained eternally and form a kind of hair around the event horizon or if such destabilized region disappears eventually, etc. We aim to answer these questions by using the more advanced double null formalism \cite{Hong:2008mw,Hwang:2010aj,Hansen:2009kn,Borkowska:2011zh}.

This paper is organized as follows: In Sec.~\ref{sec:the}, we construct the model within the context of the LVS of type IIB string theory, focusing on the potential for the volume modulus. We show that this gravitationally coupled scalar field is the lightest modulus which makes it the easiest one to destabilize. In Sec.~\ref{sec:mod}, we discuss the details of the moduli destabilization via gravitational collapses.
First, we show the details of gravitational collapses using double null numerical simulations.
Second, we discuss qualitative conditions for destabilization. Finally, in Sec.~\ref{sec:dis}, we summarize our results.

\section{\label{sec:the}The volume modulus}

\subsection{The background potential}
We work within the framework of the LARGE volume scenario of type IIB string theory \cite{Balasubramanian:2005zx}. Focusing on the bosonic sector and neglecting gauge interactions, the theory is defined by the Lagrangian
\be
\mc{L}=K_{i\bar{j}}\partial_\mu \Psi_i\partial^\mu \bar{\Psi}_j+V(\{\Psi\}),
\label{eq:L}
\ee
where $\Psi_i$ denotes a generic modulus, $K_{i\bar{j}}$ is the metric in moduli space defined by $K_{i\bar{j}}=\frac{\partial^2 K }{\partial \Psi_i \partial \bar{\Psi}_j}$.  $V(\{\Psi\})$ is the $F$-term potential given by
\begin{equation}
	V(\{\Psi\})=e^K \left(K^{i \bar{j}} D_i W D_{\bar{j}}\bar{W} -3 |W|^2 \right),
	\label{eq:V}
\end{equation}
where $D_i W=\partial_i W+W\partial_i K$.

The specification of the K\"ahler potential $K$ and of the holomorphic superpotential $W$ completely determines the action for the moduli fields. These two functions can be found explicitly via dimensional reduction of the 10 D action. It is well known that by taking only the leading terms in the perturbative expansion in the 10 D theory one ends up with a compactified theory with a no-scale structure. To see how this arises note that the K\"ahler and superpotential take the schematic form
\be
K_0=K_T(T)+K_U(U)+K_S(S)\qquad\text{and}\qquad W=W_0(U,S),
\ee
and so the scalar potential becomes
\be
V=e^{K_0} \left(K_T^{i \bar{j}} D_{Ti} W D_{\bar{T_j}}\bar{W} +{K_U}^{i \bar{j}} D_{U_i} W D_{\bar{U_j}}\bar{W}+{K_S}^{S \bar{S}} D_{S} W D_{\bar{S}}\bar{W}-3 |W|^2\right).
\ee
Non-vanishing fluxes on the compact space \cite{Giddings:2001yu}, $\langle W_0\rangle\neq0$,  stabilize the complex structure moduli ($U_i$) and the axio-dilaton $S$ at a supersymmetric locus $D_U W=D_S W=0$. These fields then get a mass at a high scale
 and can essentially be integrated out when studying the low energy physics. The following no-scale identity
 \be
{K_T}^{i \bar{j}} D_{T_i} W D_{\bar{T_j}}\bar{W}= {K_T}^{i \bar{j}} |W|^2 \partial_i K_T \partial_{\bar{j}}K_T=3 |W|^2
\ee
then implies that the K\"ahler moduli ($T_i$) survive as exactly flat directions of the potential, with all the phenomenological challenges this poses.  In particular note that at this level the theory is unable to satisfactory explain why we seem to live in 4 dimensions if spacetime is intrinsically 10 dimensional.

In order to break this structure and stabilize the K\"ahler moduli it is therefore essential to go beyond leading order and include subleading corrections to the supergravity action. In the realm of effective field theory, the corrections to the leading order action can be classified as perturbative or non-perturbative. It follows from the properties of supersymmetric field theory that the holomorphic superpotential $W$ is not renormalized and so the only new contributions to $W$ will come from non-perturbative effects. These will originate from Euclidean D3 instantons or gaugino condensation in D7 branes and generate terms of the form $W_{\mathrm{np}}\propto e^{-a_i T_i}$, such that the full superpotential for the moduli sector is given by
\begin{equation}
	W=W_0+\sum_i A_i e^{- a_i T_i}.
	\label{eq:WLVS}
\end{equation}
These non-perturbative corrections to $W$ are essential to stabilize the geometry of the compact space as initially demonstrated in \cite{Kachru:2003aw}. The K\"ahler potential is not protected by non-renormalization theorems and so it can, and in generally will, receive both perturbative and non-perturbative corrections. It is usually assumed that the perturbative contributions will be dominant. Recalling that the action is a perturbative expansion in both the string length $l_{\mathrm{s}} \equiv 2\pi \sqrt{\alpha'}$ and the string coupling $g_{\mathrm{s}}\equiv \langle \mathrm{Re}(S)\rangle$ we see that in general the perturbative K\"ahler potential can be written as
\be
K=K_0+\delta K_{g_{\mathrm{s}}}+\delta K_{\alpha'}.
\ee
Of particular relevance for the large volume constructions of \cite{Balasubramanian:2005zx} that we consider throughout this work are the  $\alpha'^3$ corrections to $K$ \cite{Becker:2002nn}. These originate from a 10 dimensional term of the form $\alpha'^ 3 \mathcal{R}^4$ and give rise to a correction to the K\"ahler potential for the K\"ahler moduli \cite{Becker:2002nn}:
\begin{equation}
	K_{\mathrm{K}}=-2 \ln \left[\mathcal{V}+\frac{\xi}{2 g_{\mathrm{s}}^{3/2}} \right],
	\label{eq:KLVS}
\end{equation}
where $\xi$ is related to the Euler number of the compact space. In the spirit of LVS compactifications we demand that $\xi>0$ \cite{Balasubramanian:2005zx}.

In order to write the scalar potential explicitly we need to specify the geometry of the compact space. We choose it to be of the Swiss-cheese type, such that its volume is written as
\be
\vo=\frac{1}{\lambda}\left[\left(\frac{T_{\mathrm{b}}+\bar{T_{\mathrm{b}}}}{2}\right)^{3/2}-\left(\frac{T_{\mathrm{s}}+\bar{T_{\mathrm{s}}}}{2}\right)^{3/2}\right]=\frac{1}{\lambda}\left(\tau_{\mathrm{b}}^{3/2}-\tau_{\mathrm{s}}^{3/2}\right),
\label{eq:Vol}
\ee
where we have used the definition $T_i\equiv \tau_i+i b_i$. Then taking into account Eqs.~(\ref{eq:WLVS}) and (\ref{eq:KLVS}), the scalar potential for the K\"ahler moduli sector can be written as:
\be
	V= \frac{8}{3}\frac{\lambda a^2 |A|^2}{\mathcal{V}}e^{-2 a \tau_{\mathrm{s}}}\sqrt{\tau_{\mathrm{s}}}-4 \frac{|A W|}{\mathcal{V}^2}a\tau_{\mathrm{s}} e^{-a \tau_{\mathrm{s}}}+\frac{3}{4}\frac{|W|^2\xi}{\mathcal{V} ^3 g_{\mathrm{s}}^{3/2}},
	\label{eq:VLVS}
\ee
in the limit where $\tau_{\mathrm{s}}\ll\tau_{\mathrm{b}} \sim\vo^{2/3}$. The position of minimum is found by solving $\frac{\partial V}{\partial \mathcal{V}}=\frac{\partial V}{\partial \tau_{\mathrm{s}}}=0$, from which we find
\begin{equation}
	\langle \mathcal{V}\rangle=\frac{3 |W_0|}{4 \lambda a |A|} \sqrt{\tau_{\mathrm{s}}} e^{ a \tau_{\mathrm{s}}}\left(1-\frac{3}{4 a \tau_{\mathrm{s}}}+\mathcal{O}\left(a\tau_{\mathrm{s}}\right)^{-2}\right),
	\label{eq:Vvev}
\end{equation}
and
\begin{equation}
	\langle\tau_{\mathrm{s}}\rangle^{3/2}\approx\frac{\lambda \xi}{g_{\mathrm{s}}^{3/2}}\left(\frac{1}{2}+\frac{1}{4 a \tau_{\mathrm{s}}}+\mathcal{O}\left(a \tau_{\mathrm{s}}\right)^{-2}\right).
	\label{eq:tausvev}
\ee
Given that $\langle\tau_{\mathrm{s}}\rangle\gsim1$ in the controllable regime of the theory, we see that the minimum for the volume naturally lies at exponentially large values: $\langle\vo\rangle\sim e^{a \tau_{\mathrm{s}}}$.

An interesting feature of the large volume minimum is the mass hierarchy in the K\"ahler moduli sector, with the volume mode substantially lighter that the small moduli. This will be important in the ensuing discussion as it allows us to integrate out the small modulus and to identify the volume mode as the easiest modulus to destabilize. To see this one computes the eigenvalues of the physical mass matrix defined as
\be
\mc{M}=\left(K^{-1}\right)_{i\bar{j}}\partial_{\bar{j}k}V
\ee
at the minimum. Noting that since $\vo\gg1$ then $1/\vo\ll1$ is a good expansion parameter. To leading order in the inverse volume expansion, the eigenvalues of $\mc{M}$ are then
\be
m_{\mathrm{b}}\sim \frac{M_{\mathrm{P}}}{\vo^{3/2}}\qquad \text{and} \qquad m_{\mathrm{s}}\sim \frac{M_{\mathrm{P}}}{\vo}.
\ee
In the large volume limit we then find $m_{\mathrm{b}} \ll m_{\mathrm{s}}$, and so at energies bellow $m_{\mathrm{s}}$ we can study single field dynamics by integrating out the heavier small modulus. Using Eq.~(\ref{eq:Vvev}) to eliminate the $\tau_{\mathrm{s}}$ dependence from Eq.~(\ref{eq:VLVS}) we find that the potential for the volume modulus can be written as
\be
V(\vo)=\frac{1}{\vo^3}\left( 1-\kappa (\log\vo)^{3/2}\right)
\label{eq:V1}
\ee
where $\alpha$ is a function of the compactification parameters that will determine the position of the minimum of the potential. In Eq.~(\ref{eq:V1}) we have neglected an unimportant overall $\mc{O}(1)$ factor.

The minimum for the volume modulus is located at
\be
\log \langle\vo\rangle \left( \sqrt{\log \langle\vo\rangle}-1/2\right)=1/\kappa
\ee
which can be approximated to $\log \langle\vo\rangle=\alpha^{-2/3}$ in the limit when $\log \langle\vo\rangle \gg 1$. As expected from the large volume scenario, the minimum is at this level AdS, its depth being given by
\be
\langle V \rangle=-\frac{\kappa}{2}\frac{ \log \langle\vo\rangle}{\langle\vo\rangle^3}.
\ee

It is essential that the LVS minimum is uplifted to Minkowsky or dS. That can be achieved by considering corrections to the potential coming from tension of anti branes at the tip of warped throats \cite{Kachru:2003aw}, $D$-terms from magnetized branes  \cite{Burgess:2003ic} or dilaton-dependent non-perturbative effects \cite{Cicoli:2012fh}. Regardless of the microscopic origin of the uplifting term it generates a term of the form
\be
V_{\mathrm{up}}=\frac{\epsilon}{\vo^p},
\ee
where $\epsilon$ is tuned such that $V(\langle\vo\rangle)=0$. Throughout this work we will assume $p=2$ as generated by $\overline{\mathrm{D3}}$ branes \cite{Kachru:2003aw}.

\subsection{The local contribution to the potential}

The last remaining contribution to the potential for the modulus is the term parameterizing the interaction with local distributions of matter. In \cite{Conlon:2010jq} it was argued that low energy mass scales and couplings depended on the volume of the compactification through RG running. In particular the dependence would arise from the fact that in string compactifications the high energy cut-off $\Lambda_{\mathrm{UV}}$ from which the couplings start running is dependent on the volume. Typically one finds that $\Lambda_{\mathrm{UV}}=M_{\mathrm{P}}/\vo^q$. The value of $q$ depends on which physical scale corresponds to  $\Lambda_{\mathrm{UV}}$. The particular value of $q$ was found to have only a limited influence on the qualitative results and so throughout this analysis we identify $\Lambda_{\mathrm{UV}}=M_{\mathrm{string}}$ which sets $q=1/2$ and yields a contribution to the modulus potential that scales as
\be
V_{\mathrm{local}}\propto\Lambda_{\mathrm{UV}}^4\propto \frac{1}{\vo^2}.
\label{eq:Vlocal}
\ee
It is convenient to formulate the problem in terms of the canonically normalized volume modulus. From Eqs.~(\ref{eq:KLVS}) and (\ref{eq:L}) we find
\be
\mc{L}_{\mathrm{K}}=\frac{3}{4 \tau_{\mathrm{b}}^2}\partial_\mu \tau_{\mathrm{b}} \partial^\mu\tau_{\mathrm{b}}
\ee
which prompts the definition
\be
\Phi\equiv\sqrt{\frac{3}{2}}\log \tau_{\mathrm{b}}=\sqrt{\frac{2}{3}}\log \vo.
\ee
The uplifted large volume potential for the canonically normalized volume modulus is then
\be
V=\left(1-\kappa \Phi^{3/2}\right)e^{-\sqrt{27/2}\Phi}+\epsilon e^{-\sqrt{6}\Phi},
\label{eq:LVSV}
\ee
and the local contribution to the potential is
\be
V_{\mathrm{local}}= \beta e^{-c\Phi} \mathcal{L}_{\mathrm{M}}.
\label{eq:vlocal}
\ee
In Fig.~\ref{fig:potential} we depict the effect of the local term, Eq.~(\ref{eq:vlocal}), on the background potential for the volume modulus, Eq.~(\ref{eq:LVSV}).

\begin{figure}[h!]
\begin{center}
\includegraphics[scale=0.75]{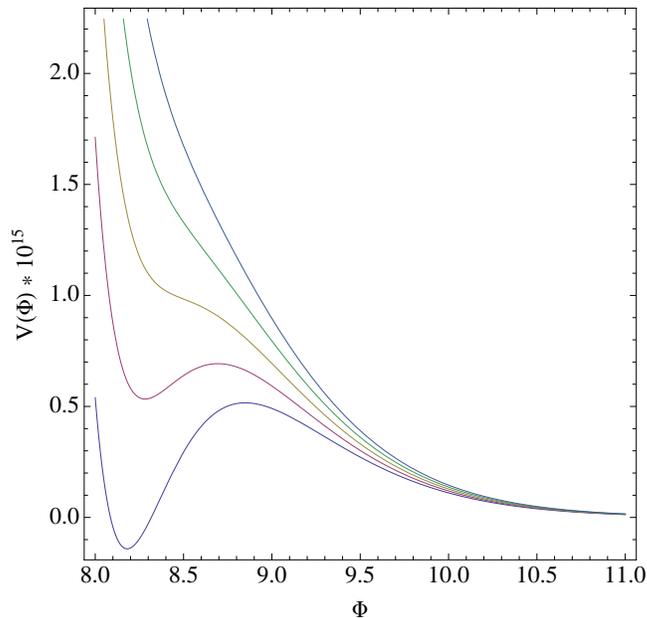}
\caption{Volume modulus potential in regions of different local density. We see that as density increases, the minimum gets lifted until it ceases to exist, leading to destabilization of the volume modulus.}
\label{fig:potential}
\end{center}
\end{figure}

\section{\label{sec:mod}Moduli destabilization and gravitational collapse}

The main aim of this paper is to study the stability of the moduli vacuum taking into account the interaction with matter. In particular we investigate if the volume modulus can be destabilized in the process of black hole formation and if an observer outside the horizon is able to probe the destabilized region.

Intuitively one expects that as an initial matter distribution collapses and the local energy density increases, the system  will eventually reach a state where the energy density is of the order of the large volume potential and is then able to destabilize the volume modulus causing a shift in its vev or in extreme cases triggering runaway and decompactification. The system's continued collapse under its own gravitational attraction eventually results in the birth of a black hole with all the matter hidden behind the event horizon. The interesting question is whether the destabilization that seems inevitable in these simple models is visible to an observer at infinity or if the destabilized region is always shielded by the event horizon.

This system was originally studied in \cite{Conlon:2010jq} where it was assumed that the black hole was formed from a initially dilute spherical distribution of pressureless dust. The spacetime inside this sphere was assumed to be a positively curved FRW which was smoothly joined to a Schwarzschild spacetime at the surface of the dust sphere. Time dependence arose only through the FRW spacetime scale factor, with the modulus assumed to lie at the local minimum of the potential. It was found that a small destabilized region would lie for a finite time outside the horizon and so an external observer would in principle be able to observe it. This region would eventually fall beyond the Schwarzschild radius making it inaccessible for outside observers.

Here we aim to extend the work of \cite{Conlon:2010jq} by considering a fully dynamical system, where both the metric and the modulus are allowed to vary over spacetime. To do so we consider a coupled system of four dimensional gravity, volume modulus and matter. The action for the system is given by
\begin{eqnarray}
S = \int dx^{4} \sqrt{-g} \left[ \frac{1}{16\pi} R - \frac{1}{2}\nabla_{\mu} \Phi \nabla^{\mu} \Phi - V(\Phi) + \beta e^{-c\Phi} \mathcal{L}_{\mathrm{M}} \right],
\end{eqnarray}
where $V(\Phi)$ is the uplifted large volume potential of Eq.~(\ref{eq:LVSV}). We model the matter component by a scalar field of mass $m$, with Lagrangian
\be
\mathcal{L}_{\mathrm{M}} = - \frac{1}{2} g^{\mu\nu} \phi_{;\mu} \phi_{;\nu} - \frac{1}{2}m^{2} \phi^{2},
\ee
and assume the spacetime metric to take the form
\be
ds^{2} = -\alpha^{2}(u,v) du dv + r^{2}(u,v) d\Omega^{2},\label{eq:uvmetric}
\ee
where $u$ and $v$ are null coordinates.

The large volume potential's parameter $\kappa$ determined the volume of the compactification and through it the moduli masses. The uplift parameter $\epsilon$ is tuned such that the vacuum at infinity is Minkowski or dS. The parameters $\beta$ and $c$ determine the strength of the interaction between the modulus and matter. Guided by the fact that in \cite{Conlon:2010jq} the value of $c$ did not have a significant impact on the results we choose $c = \sqrt{6}$. Furthermore we set $\beta \exp{-c\Phi_{\mathrm{m}}} = 1$, where $\Phi_{\mathrm{m}}$ is the local minimum of the potential $V(\Phi)$\footnote{ Note that this amounts to choosing the position of the minimum of the volume modulus $\Phi_m$.}. Note that, there is a scaling symmetry
\begin{eqnarray}
\left(\beta, \phi\right) \rightarrow \left( D\beta, \frac{\phi}{\sqrt{D}}\right)
\label{eq:scaling}
\end{eqnarray}
for arbitrary $D$. And hence, for any calculation with $\beta$, we can rescale and restore the results.

With the model in place, we consider the gravitational collapses and study the moduli destabilization process.

\subsection{\label{sec:num}The method}

In this section we solve the field equations numerically using the double null formalism. We allow for a fully dynamical metric as well as dynamic matter scalar field $\phi$ and volume modulus $\Phi$, in an interesting application of the double null formalism of scalar-tensor gravity. In \cite{Hwang:2010aj}, the authors discussed responses of the Brans-Dicke type field, but did not focus on the possibility of destabilization of the compact space.

We solve the Einstein equations:
\begin{eqnarray}\label{eq:Einstein}
G_{\mu\nu} = 8 \pi \left( T^{\Phi}_{\mu\nu} + \beta e^{-c\Phi} T^{\mathrm{M}}_{\mu\nu} \right),
\end{eqnarray}
where the stress energy tensors are
\begin{eqnarray}
\label{eq:T_Phi}
T^{\Phi}_{\mu\nu} &=& \Phi_{;\mu}{\Phi}_{;\nu} - \frac{1}{2} \Phi_{;\rho}\Phi_{;\sigma}g^{\rho\sigma}g_{\mu\nu} -V(\Phi) g_{\mu\nu},\\
\label{eq:T_M}
T^{\mathrm{M}}_{\mu\nu} &=& \phi_{;\mu}\phi_{;\nu} - \frac{1}{2} \phi_{;\rho} \phi_{;\sigma} g^{\rho\sigma} g_{\mu \nu} - \frac{1}{2}m^{2} \phi^{2} g_{\mu \nu}.
\end{eqnarray}
The field equations for the scalar fields are given by
\begin{eqnarray}
\label{eq:Phi}0 &=& \Phi_{;\mu\nu}g^{\mu\nu}- \frac{dV}{d\Phi} - c \beta e^{-c\Phi} \mathcal{L}_{\mathrm{M}}, \\
\label{eq:phi}0 &=& \phi_{;\mu\nu}g^{\mu\nu} - c\Phi_{;\mu} \phi_{;\nu} g^{\mu\nu} - m^{2} \phi.\label{eq:KG}
\end{eqnarray}

It is convenient to analyze the system in the double null coordinate system of Eq.~(\ref{eq:uvmetric}).
We define the rescaled matter field and volume modulus as:
\begin{eqnarray}
\sqrt{4\pi}\phi \equiv s, \quad \sqrt{4\pi}\Phi \equiv S,\label{eq:sandS}
\end{eqnarray}
and their derivatives with respect to the null coordinates as
\begin{eqnarray}\label{eq:conventions}
 W \equiv S_{,u},\quad Z \equiv S_{,v}, \quad w \equiv s_{,u},\quad z \equiv s_{,v}.
\end{eqnarray}
In addition, the derivatives of the metric are defined as
\bea
 g \equiv r_{,v},\quad h \equiv \frac{\alpha_{,u}}{\alpha},\quad d \equiv \frac{\alpha_{,v}}{\alpha},\quad f \equiv r_{,u}.\label{eq:ghdf}
\eea

We now write Eqs.~(\ref{eq:Einstein})-(\ref{eq:KG}) in terms of these new variables. Here we present only the final expressions, the intermediate steps are given in Appendix A. The Einstein equations are
\begin{eqnarray}
\label{eq:E1}f_{,u} &=& 2 f h - 4 \pi r T_{uu},\\
\label{eq:E2}g_{,v} &=& 2 g d - 4 \pi r T_{vv},\\
\label{eq:E3}f_{,v}=g_{,u} &=& -\frac{\alpha^{2}}{4r} - \frac{fg}{r} + 4\pi r T_{uv},\\
\label{eq:E4}h_{,v}=d_{,u} &=& -\frac{2\pi \alpha^{2}}{r^{2}}T_{\theta\theta} - \frac{f_{,v}}{r},
\end{eqnarray}
where the components of the stress-energy tensor are given by Eqs.~(\ref{eq:Tuu})-(\ref{eq:Tthth}).
The Klein-Gordon equations for the scalars become
\begin{eqnarray}
\label{eq:s}z_{,u} = w_{,v} &=& - \frac{fz}{r} - \frac{gw}{r} + \frac{c}{2 \sqrt{4 \pi}} \left( Wz +Zw \right) - \frac{1}{4} \alpha^{2} m^{2} s,\\
\label{eq:S}Z_{,u} = W_{,v} &=& - \frac{fZ}{r} - \frac{gW}{r} -\pi\alpha^{2} \left(V'(S) + \frac{c}{\sqrt{4\pi}} \beta e^{-c S/\sqrt{4\pi}} \mathcal{L}_{\mathrm{M}} \right),
\end{eqnarray}
where the matter Lagrangian is
\begin{eqnarray}
\mathcal{L}_{\mathrm{M}} = \frac{wz}{2\pi\alpha^{2}} - \frac{m^{2}}{8\pi}s^{2}.
\end{eqnarray}

The physics of the interplay between gravitational collapse and moduli stability is encoded by the solutions of the set of coupled first order differential Eqs.~(\ref{eq:E1})-(\ref{eq:S}), for which me must provide appropriate initial conditions.

\subsection{Initial conditions}

We need initial conditions for all functions ($\alpha, h, d, r, f, g, S, W, Z, s, w, z$) on the initial $u=u_{\mathrm{i}}$ and $v=v_{\mathrm{i}}$ surfaces, where we set $u_{\mathrm{i}}=v_{\mathrm{i}}=0$.

We have gauge freedom to choose the initial $r$ function. Although all constant $u$ and $v$ lines are null, there remains freedom to choose the distances between these null lines. Here, we choose $r(0,0)=r_{0}$, $f(u,0)=r_{u0}$, and $g(0,v)=r_{v0}$, where $r_{u0}<0$ and $r_{v0}>0$ such that the radial function for an in-going observer decreases and that for an out-going observer increases.

\begin{description}
\item[In-going null surface:]
We use a shell-shaped scalar field. Therefore, its interior is not affected by the shell. First, it is convenient to choose $r_{u0}=-1/2$ and $r_{v0}=1/2$; we choose that the mass function on $u_{\mathrm{i}}=v_{\mathrm{i}}=0$ vanish, where the Misner-Sharp mass is
\begin{eqnarray}
m(u,v) = \frac{r}{2} \left( 1 + \frac{4 r_{,u} r_{,v}}{\alpha^{2}} - \frac{8\pi V(S)}{3} r^{2} \right).
\end{eqnarray}
Hence, to specify a pure de Sitter background, for given $r(0,0)=r_{0}$ and $S(0,0)=S_{\mathrm{m}}$ (local minimum), then
\begin{eqnarray}
\alpha(0,0) = \left( 1 - \frac{8 \pi V(S_{\mathrm{m}})}{3} \right)^{-1/2}.
\end{eqnarray}
In addition, $S(u,0)=S_{\mathrm{m}}$ and $W(u,0)=s(u,0)=w(u,0)=h(u,0)=0$ hold.

We need more information to determine $d, g, z$, and $Z$ on the $v=0$ surface. We obtain $d$ from Eq.~(\ref{eq:E4}), $g$ from Eq.~(\ref{eq:E3}), $z$ from Eq.~(\ref{eq:s}), and $Z$ from Eq.~(\ref{eq:S}).

\item[Out-going null surface:]
We first choose $S(0,v)=S_{\mathrm{m}}$. We can choose an arbitrary function for $s(0,v)$ to induce a collapsing pulse. In this paper, we use
\begin{eqnarray}
s(u_{\mathrm{i}},v)= A\sqrt{2D} \sin^{2} \left( \pi \frac{v-v_{\mathrm{i}}}{v_{\mathrm{f}}-v_{\mathrm{i}}} \right) \cos \left( 2 \pi \frac{v-v_{\mathrm{i}}}{v_{\mathrm{f}}-v_{\mathrm{i}}} \right)
\end{eqnarray}
for $v_{\mathrm{i}}\leq v \leq v_{\mathrm{f}}$ and otherwise $s(u_{\mathrm{i}},v)=0$, where $u_{\mathrm{i}}=0$, $v_{\mathrm{i}}=0$, and $v_{\mathrm{f}}=20$ denotes the end of the pulse in the initial surface. We then obtain $z(0,v)=s(0,v)_{,v}$. This implements one pulse of energy ($T_{vv} \sim z^{2}$) along the out-going null direction by the continuous function $z(0,v)$.

Furthermore, from Eq.~(\ref{eq:E2}) we can obtain $d(0,v)$, since $g_{,v}(0,v)=0$. By integrating $d$ along $v$, we get $\alpha(0,v)$.

We need more information for $h, f, w$ and $W$ on the $u=0$ surface. We obtain $h$ from Eq.~(\ref{eq:E4}), $f$ from Eq.~(\ref{eq:E3}), $w$ from Eq.~(\ref{eq:s}), and $W$ from Eq.~(\ref{eq:S}). This finishes the assignments of the initial conditions.
\end{description}

Finally, we can interpret this setup as follows (Fig.~\ref{fig:domain}). We obtain a numerical result for a given integration domain $(u=0,u=u_{\mathrm{max}}) \times (v=0, v=v_{\mathrm{max}})$ (left). By tilting $45$-degree, we obtain a Penrose diagram (middle), since the two coordinates are null. Initially, there was no black hole, as the matter shell collapses a black hole forms. In the distant future, the geometry asymptotically approaches  that of a static neutral black hole  (right).

\begin{figure}
\begin{center}
\includegraphics[scale=0.85]{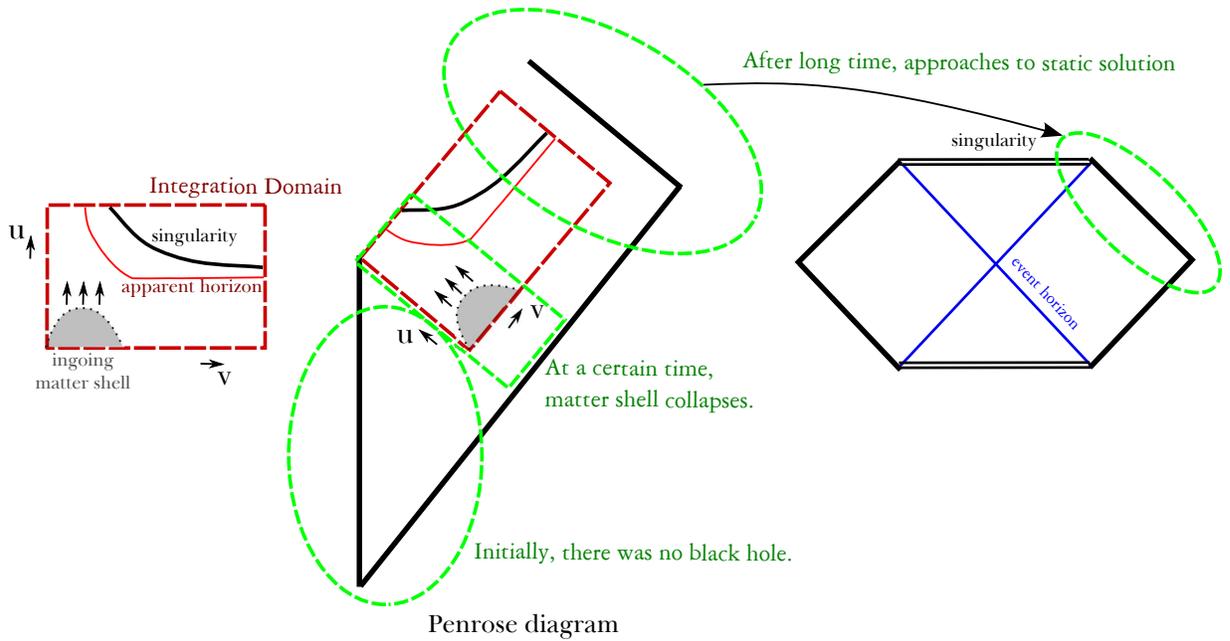}
\caption{\label{fig:domain}Left: We obtain the left figure from simulations. Middle: Tilting $45$-degree, we obtain a Penrose diagram. Initially, there was no black hole. At a certain time, a matter shell collapses and a black hole is generated. After a long time, the geometry approaches a static limit. Right: The Penrose diagram for a static neutral black hole.}
\end{center}
\end{figure}

\subsection{Simulations and results}

We run the simulations with the aim of testing the intuitive picture developed in \cite{Conlon:2010jq}, namely that the denser the initial matter distribution, the more drastic the destabilization will be. The two relevant parameters to vary in this context are the mass and the amplitude of the matter scalar field, $m$ and $A$ respectively. We also want to probe how the height of the modulus potential barrier influences the dynamics and the final state of the system. This can be done by varying $\kappa$ keeping the remaining parameters unchanged.

With this in mind we perform 3 distinct runs:
\begin{itemize}
\item Run 1: We keep  $A=7000$, $\kappa=0.05$ fixed and vary the mass of the matter field in the range $m^{2}=0.01, 0.05, 0.1, 0.2$;
\item Run 2: We keep   $\kappa = 0.05$ , $m^2 = 0.05$ fixed and vary the amplitude of the matter field in the range $A = 6000, 6500, 7000, 7500$;
\item Run 3: We keep   $A = 7000$ , $m^2 = 0.05$ fixed and vary the modulus potential parameter in the range $\kappa = 0.05, 0.050001, 0.05001, 0.0501$, adjusting $\epsilon$ accordingly.
\end{itemize}

Throughout all 3 runs we keep the initial size of the matter shell fixed at $r_{0} = 10$ and its range at $v_{\mathrm{f}}=20$. Furthermore we can use the scaling freedom of Eq.~(\ref{eq:scaling}) to set $\beta = \exp{c\Phi_{\mathrm{m}}}$ and set the parameter in the exponential $c=\sqrt{6}$, as in \cite{Conlon:2010jq}.

For any given choice of parameters we then can calculate all functions in the integration domain, using the second order Runge-Kutta method \cite{nr}.  We have checked the convergence and consistency of the simulation and present the analysis in Appendix B.

\begin{figure}
\begin{center}
\includegraphics[scale=0.27]{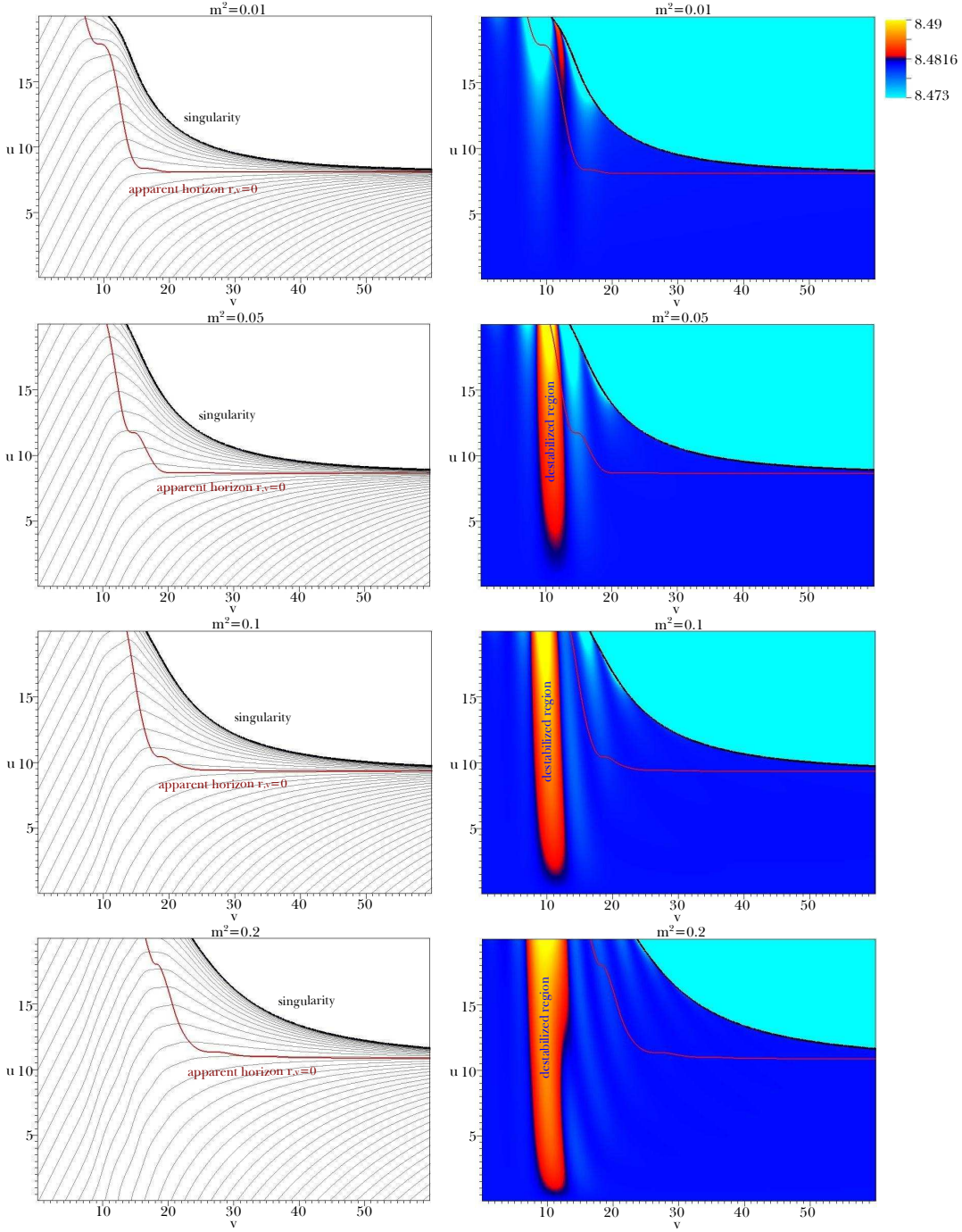}
\caption{Run 1 results: solutions for $r$ and $\Phi$ for $m^{2}=0.01, 0.05, 0.1, 0.2$, $A=7000$, $\kappa=0.05$.}
\label{fig:plots}
\end{center}
\end{figure}

\begin{figure}
\begin{center}
\includegraphics[scale=1]{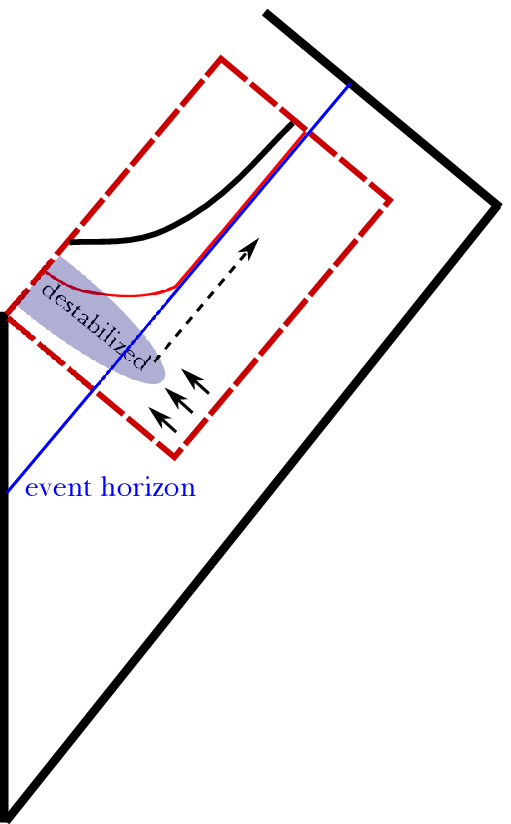}
\caption{\label{fig:interpretation}Interpretation of the result on the Penrose diagram. Some effects of the destabilized region (dotted arrow) can be observed by an asymptotic observer, since the destabilized region can be outside of the event horizon (blue line).}
\end{center}
\end{figure}


In Fig.~\ref{fig:plots}, we plot the result of Run 1, where we test the effect of the mass of the matter field on the stability of the volume modulus. The plots on the left show the lines of constant radius while the plots on the right display the profile of the volume modulus. We start by observing that all causal structures show a formation of a typical neutral black hole: a singularity is space-like and an apparent horizon is also space-like and approaches to a null direction. However, the dynamics of modulus field $\Phi$ is non trivial: we see that there are regions in spacetime where $\Phi$ stays at its background minimum (sky blue-blue region in Fig.~\ref{fig:plots}) but where the local density is high enough, the volume modulus moves beyond the position of the local maximum (yellow-red region in Fig.~\ref{fig:plots}). As the mass $m^{2}$ of the matter field increases, the destabilized region grows. The increased destabilized region postpones the formation of the black hole. However, such non-trivial field dynamics eventually disappears as sufficient time elapses, as one would expect from the \textit{no-hair theorem}. One crucial point is that the destabilized region is partially outside of the event horizon. In principle any physical process happening in that region will differ from the same process taking place at infinity, since the different volume modulus vacuum expectation value can lead to different masses and couplings. Outgoing light from that region (that in Fig.~\ref{fig:plots} travels along horizontal lines) can reach an asymptotic observer sitting at infinity (see Fig.~\ref{fig:interpretation} for an interpretation). This observer is therefore able to probe a region of spacetime where the standard model masses and couplings are distinct from the ones measured in the laboratory. To conclude the analysis of Run 1, we note that the destabilization is more severe for larger values of the matter field's mass, as one intuitively expected.

\begin{figure}
\begin{center}
\includegraphics[scale=0.27]{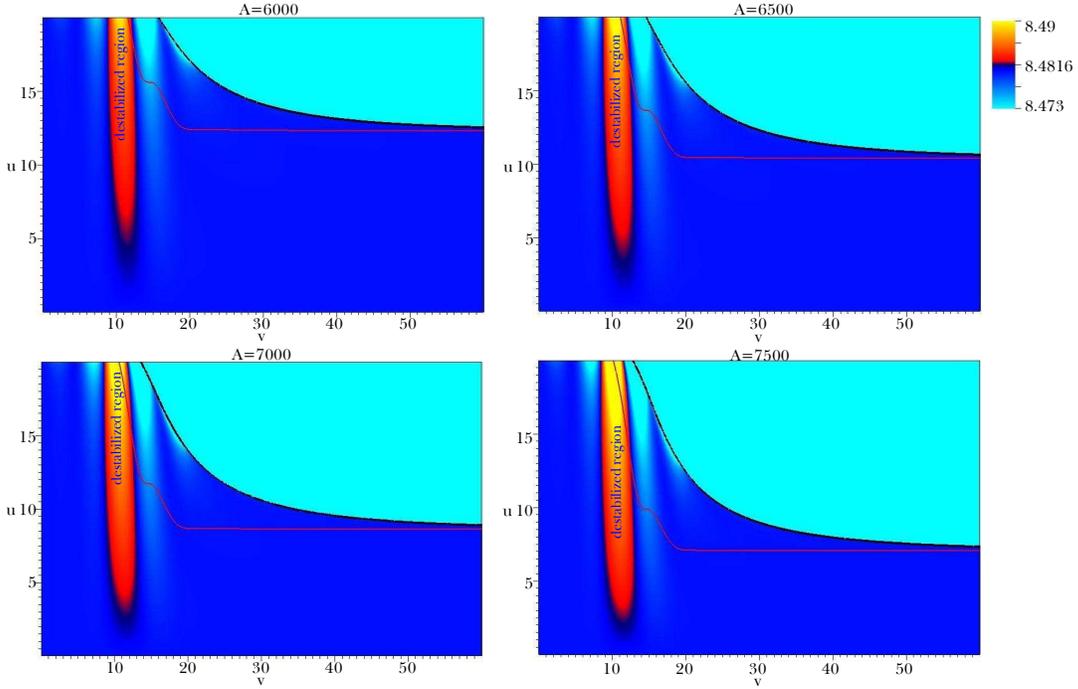}
\caption{Run 2 results: $\Phi$ profile for $A=6000, 6500, 7000, 7500$, $\kappa=0.05$, $m^{2}=0.05$.}
\label{fig:VaryingAmplitude}
\end{center}
\end{figure}
\begin{figure}
\begin{center}
\includegraphics[scale=0.27]{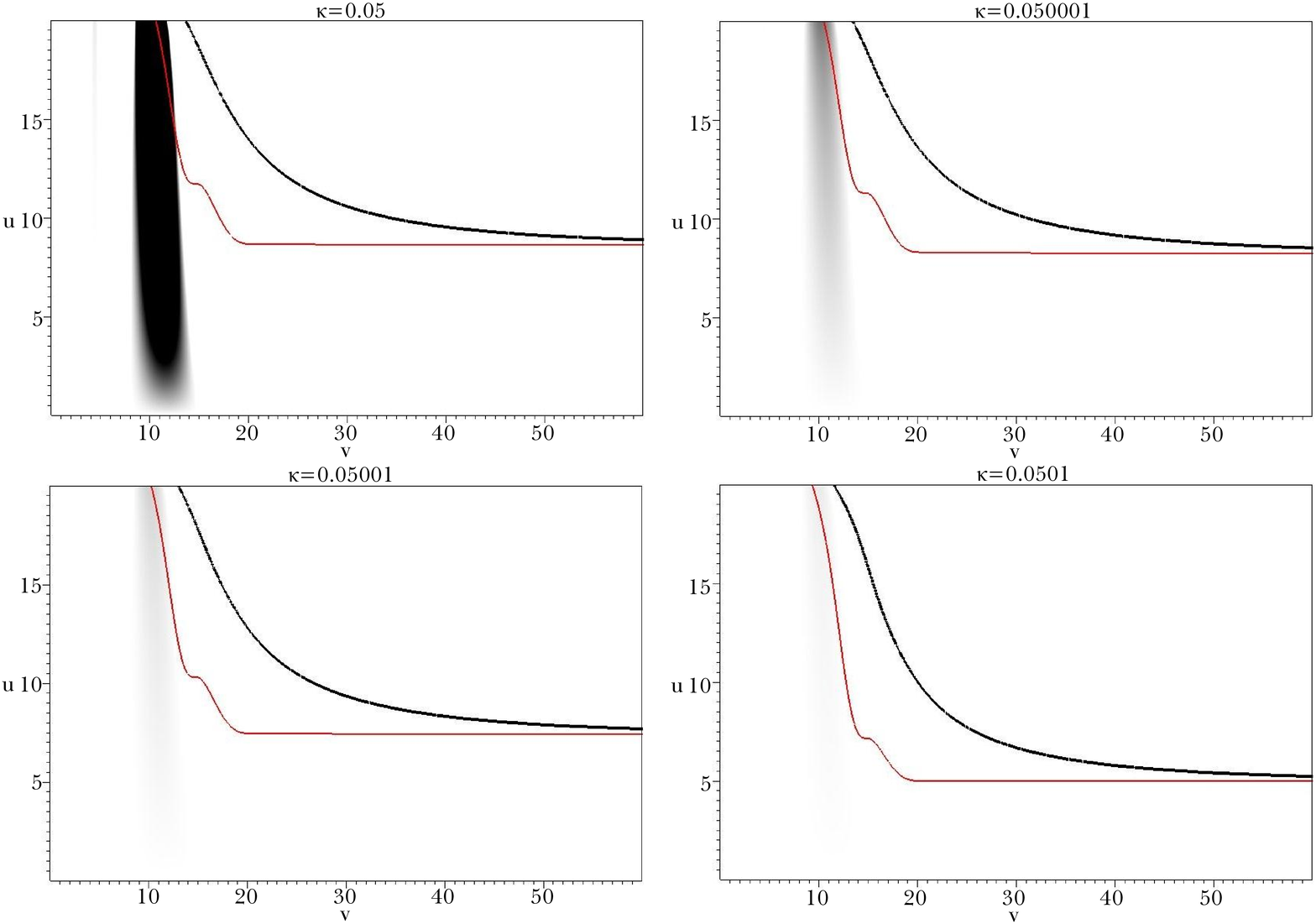}
\caption{Run 3 results: $\Phi$ profile for $\kappa=0.05, 0.050001, 0.05001, 0.0501$, $A=7000$, $m^{2}=0.05$. The white region corresponds to the local minimum and the black region denotes the field value greater than the local maximum.}
\label{fig:VaryingPotential}
\end{center}
\end{figure}


In Run 2, whose results are presented in Fig.~\ref{fig:VaryingAmplitude}, we vary the amplitude of the matter field, $A$. The results are similar to Run 1 and in accordance with the expectation that the larger the amplitude, the more energy will be stored in the matter shell  and the more pronounced the destabilization of the volume modulus. In addition, as the field amplitude increases, the size of the event horizon also increases. 

Fig.~\ref{fig:VaryingPotential} depicts the effects of varying $\kappa$, Run 3. The change of $\kappa$ implies the change of the mass scale around the local minimum of the volume modulus's potential and the change in the height of the potential barrier separating the minimum from decompactification. We note that the potential is very sensitive to the value of $\kappa$ and so it suffices to vary this quantity in a very narrow range. Here the black region corresponds to $\Phi$ beyond the local maximum and white region to the vicinity of the local minimum. As $\kappa$ increases, the black coloured region suddenly disappears. In other words, as we increase $\kappa$, the mass scale around the local minimum increases, and hence the moduli field is confined by the local minimum.

\subsection{Conditions for destabilization}

Having seen from the numerical results that destabilization of the volume modulus within a very dense region is indeed possible, we now try to identify the conditions for such behaviour. The fundamental premise of this work is that in the presence of the matter distribution the moduli potential gets modified to 
\begin{eqnarray}
V_{\mathrm{eff}}(\Phi) =\underbrace{\left(1-\kappa \Phi^{3/2} \right) e^{-\sqrt{\frac{27}{2}}\Phi} + \epsilon e^{-\sqrt{6}\Phi}}_{\equiv V} + \mathcal{L}_{\mathrm{M}} e^{-\sqrt{6}\Phi}.
\end{eqnarray}

The destabilization will begin when the effective potential $V_{\mathrm{eff}}$  ceases to have a local minimum. To first approximation this happens when both extrema (LVS minimum and the potential barrier's maximum) become degenerate and give rise to a saddle point. We define the critical correction term $\Delta \epsilon$ such that
\begin{eqnarray}
V_{\mathrm{eff}}(\Phi) &=& \left(1-\kappa \Phi^{3/2} \right) e^{-\sqrt{\frac{27}{2}}\Phi} + \epsilon e^{-\sqrt{6}\Phi} + \Delta \epsilon e^{-\sqrt{6}\Phi},
\end{eqnarray}
and this correction term makes
\begin{eqnarray}
V_{\mathrm{eff}}'(\Phi'_{\mathrm{m}}) = V_{\mathrm{eff}}'(\Phi'_{\mathrm{M}})=0
\qquad\text{and}\qquad
\left| \Phi'_{\mathrm{m}} - \Phi'_{\mathrm{M}} \right| = 0.
\end{eqnarray}
Clearly, $\Delta \epsilon$ depends on the moduli potential shape. If $\mathcal{L}_{\mathrm{M}} \gtrsim \Delta \epsilon$, then the moduli field can start to roll.

This condition on the local energy density is a necessary condition but it is not sufficient to guarantee destabilization in such a  highly dynamical process. If the vacuum energy dominated region is too short (in time or length scale), then the modulus field will not roll sufficiently and hence it will be perturbed but not destabilized. To guarantee that destabilization will take place we require the width $\Delta v$ of the matter/vacuum energy dominated region to be sufficiently wide. From the modulus field equation of motion, Eq.~(\ref{eq:S}), if the gradients of the scalar field $\Phi_{,u}\sim W $ and $\Phi_{,v}\sim Z $ are sufficiently small (and hence the vacuum energy is dominant), then
\begin{eqnarray}
S_{,uv} \simeq - \pi \alpha^{2} V_{\mathrm{eff}}(S)',
\end{eqnarray}
and the field moves $\Delta S$ after the time scale $\Delta u$ and $\Delta v$
\begin{eqnarray}
\Delta S \simeq \pi \alpha^{2} V_{\mathrm{eff}}(S)' \Delta u \Delta v.
\end{eqnarray}
This is a crude but qualitatively good approximation when the scalar field monotonely increases. For example, Fig.~\ref{fig:appS} is a part of $u=0.15$ slice from the simulation of the $m^{2} = 0.2$ case in Run 1. Here, we compared the correct result of $\Phi$ and the approximated result $\Phi_{\mathrm{approx}} = \pi \times V_{\mathrm{eff}}' \times \Delta u$. The proportionality holds as long as the field monotonely increases. Of course, as $u$ increases and as $v$ also increases, such a naive approximation is not so good. However, this can be used to estimate the maximum value that the moduli field can reach.

\begin{figure}
\begin{center}
\includegraphics[scale=0.35]{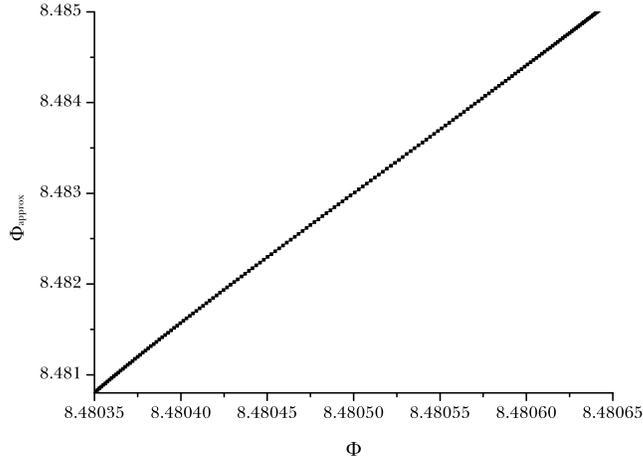}
\caption{Comparison with the $\Phi$ and approximated $\Phi_{\mathrm{approx}}$ on $u=0.15$ surface, for $m^{2}=0.2$ in Run 1.}
\label{fig:appS}
\end{center}
\end{figure}

We can reasonably assume that spacetime around that region is only moderately curved, which allows us to choose $\alpha \sim 1$, and that $\Delta u \sim \Delta v$ with the similar time/length scale. If the field moves a distance $\Delta \Phi$ from the local minimum to the local maximum of the {\it original} potential, we require the reasonable time/length scale of the vacuum energy dominated region:
\begin{eqnarray}\label{eq:condition}
\Delta v \simeq \sqrt{\frac{4 \Delta \Phi}{V_{\mathrm{eff}}(\Phi)'}}.
\end{eqnarray}

If the vacuum energy dominated region is sufficiently wide, of the order of $\Delta v$, then one expects the modulus field to be destabilized beyond the local maximum of the original potential. However, note that the process is highly dynamical and hence detailed observation of numerical calculations is crucial.

\section{\label{sec:dis}Discussion}

In this work we have investigated the interplay between gravitational collapse and moduli stability in the process of black hole formation. We have worked within the framework of the large volume scenario where the lightest modulus, and therefore the most easily destabilized, corresponds to the scalar field parameterizing the volume of the compact space. Modeling the black hole formation with a collapsing scalar field shell we have established that the volume modulus can indeed be destabilized and that the effects of this destabilization are visible to an asymptotic observer at infinity. The fact that, for a finite period of time, the destabilized region is accessible from the outside of the black hole is a rather interesting since it opens a window to regions where the physical couplings are different from the ones measured in less extreme environments. The fundamental reason behind this observation is that in the context of string theory mass scales and couplings are given as functions of the moduli vevs. If these vevs change, as they do in the destabilized region, masses and couplings will change too. As an example one can imagine that if an electron-positron pair annihilate in the destabilized region, the resulting photons will have energies different from the $511$~KeV one would expect if the same process happened away from the black hole. Since we have established that these final state photons can travel towards an observer at infinity, this observer would have access to a spacetime region with different laws of physics. 

We have shown that two conditions must be met in order to ensure destabilization of the volume modulus: the local energy density must be sufficiently high and the thickness of the vacuum energy dominated region must be wide enough.

It is worth noting that traditionally there were two known ways to destabilize a scalar field in a potential. Firstly, it is possible to destabilize via various quantum tunneling channels \cite{Coleman:1980aw}. Secondly, it is also possible that a field can be classically destabilized via bubble collisions \cite{Johnson:2011wt}. In this paper, we establish that there is a third way: \textit{a field can be destabilized by gravitational collapse, when there is a non-minimal coupling between the field and gravity}. 

The results presented here assume one particular moduli stabilization mechanism and one explicit form of the matter/moduli coupling, generalization to other models of moduli stabilization, to other types of non-minimal coupling, and to other kinds of matter fields remain interesting open problems. In addition, if such a destabilization is possible via gravitational collapses, then interesting physics could in principle also be found in bubble collisions. For this case, the destabilized region can expand to the asymptotic region, giving rise to fully fledged decompactification. This would be a more extreme final state than the one found in this study where the extent of the destabilized region was limited. We plan to address this in future work.

\subsection*{Acknowledgments}
We grateful to Joe Conlon for collaboration in the initial stages of this project.
FGP would also like to thank Alexander Westphal for interesting discussions and the University of Oxford, where this project was initiated. DY and DH are supported by the National Research Foundation of Korea (NRF) grant funded by the Korea government (MEST) through the Center for Quantum Spacetime (CQUeST) of Sogang University with grant number 2005-0049409. DY is supported by the JSPS Grant-in-Aid for Scientific Research (A) No.~21244033. DH is supported by Korea Research Foundation grants (KRF-313-2007-C00164, KRF-341-2007-C00010) funded by the Korean government (MOEHRD) and BK21.

\begin{appendix}

\begin{figure}
\begin{center}
\includegraphics[scale=1]{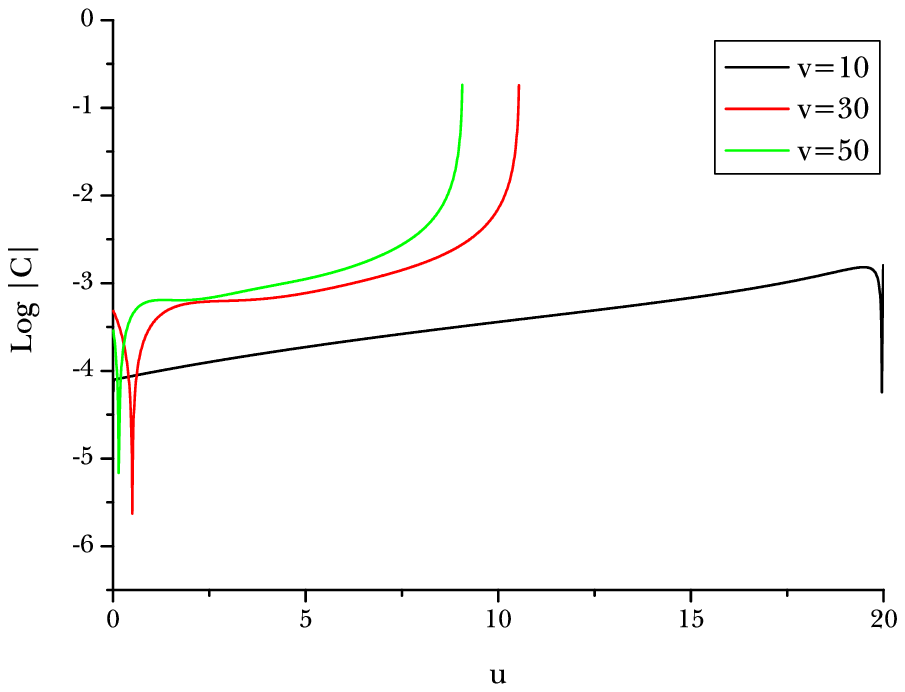}
\caption{\label{fig:constraint}Constraint function $C$ for $\kappa=0.05$, $A=7000$, $m^{2}=0.05$.}
\includegraphics[scale=1]{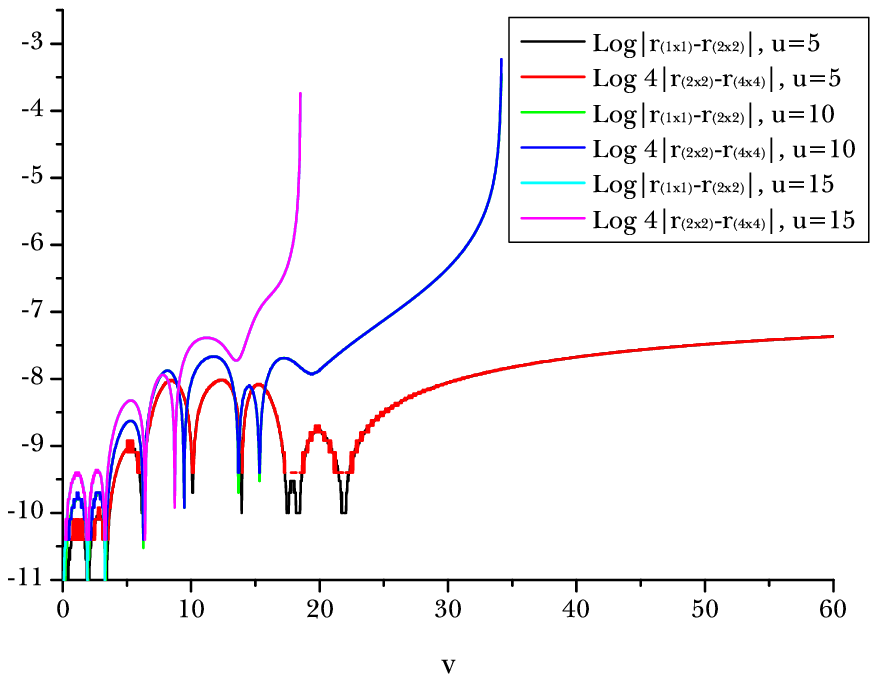}
\caption{\label{fig:convergence}Convergence test for $\kappa=0.05$, $A=7000$, $m^{2}=0.05$.}
\end{center}
\end{figure}

\section{Einstein and stress-energy tensor components}

In this appendix we compute the components of the Einstein tensor and of the stress-energy tensor in terms of the variables defined in Eqs.~(\ref{eq:sandS})-(\ref{eq:ghdf}).
The components of the Einstein tensor are:
\begin{eqnarray}
\label{eq:Guu}G_{uu} &=& -\frac{2}{r} \left(f_{,u}-2fh \right),\\
\label{eq:Guv}G_{uv} &=& \frac{1}{2r^{2}} \left( 4 rf_{,v} + \alpha^{2} + 4fg \right),\\
\label{eq:Gvv}G_{vv} &=& -\frac{2}{r} \left(g_{,v}-2gd \right),\\
\label{eq:Gthth}G_{\theta\theta} &=& -4\frac{r^{2}}{\alpha^{2}} \left(d_{,u}+\frac{f_{,v}}{r}\right).
\end{eqnarray}
The components of the energy-momentum tensor are:
\begin{eqnarray}
\label{eq:Tuu}8\pi T_{uu} &=& 8\pi\left(T^{\Phi}_{uu} + \beta e^{-cS/\sqrt{4\pi}}T^{\mathrm{M}}_{uu} \right),\\
\label{eq:Tuv}8\pi T_{uv} &=& 8\pi\left(T^{\Phi}_{uv} + \beta e^{-cS/\sqrt{4\pi}}T^{\mathrm{M}}_{uv} \right),\\
\label{eq:Tvv}8\pi T_{vv} &=& 8\pi\left(T^{\Phi}_{vv} + \beta e^{-cS/\sqrt{4\pi}}T^{\mathrm{M}}_{vv} \right),\\
\label{eq:Tthth}8\pi T_{\theta\theta} &=& 8\pi\left(T^{\Phi}_{\theta\theta} + \beta e^{-cS/\sqrt{4\pi}}T^{\mathrm{M}}_{\theta\theta} \right),
\end{eqnarray}
where
\begin{eqnarray}
\label{eq:TPhiuu}T^{\Phi}_{uu} &=& \frac{1}{4 \pi} W^{2},\\
\label{eq:TPhiuv}T^{\Phi}_{uv} &=& \frac{\alpha^{2}}{2} V(S),\\
\label{eq:TPhivv}T^{\Phi}_{vv} &=& \frac{1}{4 \pi} Z^{2},\\
\label{eq:TPhithth}T^{\Phi}_{\theta\theta} &=& \frac{r^{2}}{2 \pi \alpha^{2}} WZ -r^{2}V(S),
\end{eqnarray}
and
\begin{eqnarray}
\label{eq:TMuu}T^{\mathrm{M}}_{uu} &=& \frac{1}{4\pi} w^{2},\\
\label{eq:TMuv}T^{\mathrm{M}}_{uv} &=& \frac{\alpha^{2}}{16\pi}m^{2} s^{2},\\
\label{eq:TMvv}T^{\mathrm{M}}_{vv} &=& \frac{1}{4\pi} z^{2},\\
\label{eq:TMthth}T^{\mathrm{M}}_{\theta\theta} &=& \frac{r^{2}}{2\pi\alpha^{2}} wz - \frac{r^{2}}{8\pi}m^{2} s^{2}.
\end{eqnarray}

\section{Consistency and convergence checks}

In this appendix, we report on the consistency and convergence tests for our simulations. As a demonstration, we check the case $\kappa=0.05$, $A=7000$, $m^{2}=0.05$.

For consistency, we test one of the constraint functions:
\begin{equation}
C = \frac{\left|f_{,u}-2fh+4\pi r T_{uu}\right|}{\left|f_{,u}\right|+\left|2fh\right|+\left|4\pi r T_{uu}\right|}
\end{equation}
around $v=10, 30, 50$. Fig.~\ref{fig:constraint} shows that it is less than $1$~\% except some points, where the denominator oscillatory vanishes ($f_{,u}\approx 0$). This will not be accumulated as one integrates along $v$. Therefore, this shows good consistency.

For convergence, we compared finer simulations: $1\times1$, $2\times2$, and $4\times4$ times finer for around $u=5, 10, 15$.
In Fig.~\ref{fig:convergence}, we see that the difference between the $1\times1$ and $2\times2$ times finer cases is $4$ times the difference between the $2\times2$ and $4\times4$ times finer cases,
and thus our simulation converges to second order. The numerical error is $\lesssim 10^{-5}\%$, except near the singularity.

\end{appendix}

\newpage

\bibliographystyle{JHEP}

\end{document}